\documentclass[prd,twocolumn,showpacs,superscriptaddress,nofootinbib,floatfix,showkeys,10pt]{revtex4-2}

\usepackage{graphicx}
\usepackage{amsmath}
\usepackage{bm}
\usepackage{yhmath}
\usepackage{mathtools}
\usepackage{wasysym}
\usepackage[colorlinks,citecolor=blue,urlcolor=blue,linkcolor=blue]{hyperref}
\usepackage{subfigure}
\usepackage{color}
\usepackage{cases}
\usepackage{subfigure}
\usepackage{times}
\usepackage{dcolumn,booktabs,bm}
\usepackage{slashed}
\usepackage{amsfonts,amssymb,stmaryrd,latexsym,amsmath}
\usepackage{textcomp}
\usepackage{multirow}
\usepackage{cancel}
\usepackage{array}
\usepackage{orcidlink}
\usepackage{enumitem}
\usepackage{dashrule}

\allowdisplaybreaks[1]

 \newcommand{\clabel}[2][]{#2}
 \newcommand{\change}[1]{#1}
\newcommand{\etal}{\textit{et al}. }
\renewcommand{\arraystretch}{1.8}
\begin{document}

%%%%%%%%%%%%%%%%open the reply mode%%%%%%%%%%%%%%%%%%
%\onecolumngrid
%\input{reply1.tex}
%\newpage
%\setcounter{page}{0}
%%%%%%%%%%%%%%%%open the reply mode%%%%%%%%%%%%%%%%%%

\title{The ${\phi NN,J/\psi NN,\eta_c NN}$ systems based on HAL QCD interactions}

\author{Liang-Zhen Wen\,\orcidlink{0009-0006-8266-5840}}
%\email{wenlzh\_hep-th@stu.pku.edu.cn}
\affiliation{School of Physics, Peking University, Beijing 100871, China}

\author{Yao Ma\,\orcidlink{0000-0002-5868-1166}}\email{yaoma@pku.edu.cn}
\affiliation{School of Physics and Center of High Energy Physics, Peking University, Beijing 100871, China}

\author{Lu Meng\,\orcidlink{0000-0001-9791-7138}}\email{lu.meng@rub.de}
\affiliation{Institut f\"ur Theoretische Physik II, Ruhr-Universit\"at Bochum,  D-44780 Bochum, Germany }

\author{Shi-Lin Zhu\,\orcidlink{0000-0002-4055-6906}}\email{zhusl@pku.edu.cn}
\affiliation{School of Physics and Center of High Energy Physics, Peking University, Beijing 100871, China}

\begin{abstract}

We investigate the existence of bound states and resonances in the ${\phi NN, J/\psi NN, \eta_c NN}$ systems using HAL QCD interactions for ${\phi N, J/\psi N}$, and ${\eta_c N}$. We employ the Gaussian expansion method to solve the complex-scaled Schrödinger equation and find no resonances or bound states in the ${J/\psi NN}$ and ${\eta_c NN}$ systems. We estimate the interaction between charmonium and nuclei, concluding that the $J/\psi$ or $\eta_c$ is likely to bind with ${}^3\mathrm{H}$, ${}^3\mathrm{He}$, ${}^4\mathrm{He}$, and heavier nuclei. For the $\phi NN$ system, the lattice QCD $\phi N\left({ }^2 S_{1 / 2}\right)$ interaction is absent. We combine the  $\phi p$ correlation function analysis and HAL QCD results in Model A. We assume the spin-spin interactions for  $J/\psi N$ and $\phi N$ systems are inversely proportional to their masses in Model B. Model A predicts a stronger $\phi N({}^2 S_{1/2})$ interaction and permits a two-body bound state, whereas Model B suggests the interaction is attractive but too weak to form a bound state. Both models predict bound states for the $I(J^P) = 0(0^-)$ and $0(1^-)$ $\phi NN$ systems. In Model A, these states are deeply bound with binding energies exceeding 15 MeV and remain existent when considering parameter uncertainties. In contrast, these states are very loosely bound in Model B, with binding energies below 1 MeV and an existent probability of about 60\% when parameter uncertainties are considered. In both models, there exist very loosely bound $I(J^P) = 0(2^-)$ three-body states which resemble a $\phi$-d atom with the $\phi$ meson surrounding the deuteron, but their existences are sensitive to parameter uncertainties. No bound states or resonances are found in the isovector $I(J^P) = 1(1^-)$ $\phi NN$ system.

\end{abstract}

\maketitle

\section{Introduction}~\label{sec:intro}

The fundamental theory giving rise to strong interaction among hadrons is Quantum Chromodynamics (QCD). However, due to the nonperturbative properties of QCD at low energy scale, many issues about hadronic interaction remain unresolved to this day. Over the past several decades, the nuclear force has been the most extensively studied hadronic interaction~\cite{Machleidt:1987hj,Epelbaum:2008ga,Machleidt:2011zz}, primarily driven by the light meson exchange. 
However, this mechanism is inapplicable to interactions involving heavy quarkonium, since heavy quarkonium and nucleons do not share the same valence light quarks. Therefore, exploring the interactions between charmonium and nucleons could provide insights into different aspects of QCD that cannot be revealed through the study of nucleon-nucleon (\(NN\)) systems. Recently, the discovery of pentaquark states in the $J/\psi p$ final states~\cite{LHCb:2015yax, LHCb:2019kea} has renewed interest in the $J/\psi N$ interaction, as it plays a key role in the coupled-channel dynamics underlying the formation and production of pentaquarks~\cite{Wang:2019ato,Wang:2019nvm,Meng:2022ozq}, see Refs.~\cite{Chen:2016qju,Esposito:2016noz,Guo:2017jvc,Liu:2019zoy,Brambilla:2019esw,Chen:2022asf,Liu:2024uxn,Meng:2022ozq,Wang:2025sic} for more comprehensive reviews.  Another related system is the \(\phi N\), where the \(\phi\) meson is an almost pure \(s\bar{s}\) state. The significant mass hierarchy between the strange quark and the \(u/d\) quarks occasionally makes it resemble a heavy quark~\cite{Wang:2023hpp}. Understanding the $\phi N$ interaction is also crucial for exploring hidden strange pentaquarks~\cite{Marse-Valera:2023fnv}.

Several approaches have been used to describe the interactions between quarkonium and nucleons: QCD van der Waals 
forces~\cite{Appelquist:1978rt,Peskin:1979va}, and interactions allowed by the OZI rule, which arise from the non-zero components of $c\bar{c},s\bar{s}$ in the nucleon~\cite{Ellis:1988jwa}, as well as meson–baryon coupled channel dynamics~\cite{Cabrera:2017agk,Cabrera:2016rnc,Klingl:1997tm}. The possibility of quarkonium-nucleus bound states has also been investigated. In the 1980s, Brodsky, Schmidt, and de Teramond proposed that nuclear matter could bind quarkonium through the QCD van der Waals force, which becomes dominant when two interacting color-singlet hadrons share no common quarks~\cite{Brodsky:1989jd}. Building on this idea, Gao \etal~\cite{Gao:2000az} suggested the potential existence of a \(\phi N\) bound state, as the interaction is expected to be enhanced by a factor of \((\frac{m_c}{m_s})^3\) when transitioning from \(c\bar{c}\) to \(s\bar{s}\). This proposition sparked significant theoretical interest, leading to extensive studies on nucleus-bound quarkonium~\cite{Huang:2005gw,Wu:2012wta,Hiyama:2003cu,Cobos-Martinez:2017woo,Cobos-Martinez:2020ynh,Cobos-Martinez:2017vtr,Luke:1992tm,Kharzeev:1994pz,Wasson:1991fb,Krein:2017usp,Beane:2014sda,Alberti:2016dru,Voloshin:2007dx,Lee:2000csl,Kaidalov:1992hd,Brodsky:1997gh,deTeramond:1997ny,Sibirtsev:2005ex,Klingl:1998sr,Hayashigaki:1998ey,Kim:2000kj,Kumar:2010hs,Krein:2010vp,Tsushima:2011kh,Tsushima:2011fg,Filikhin:2024xkb,Cobos-Martinez:2023hbp,Metag:2017yuh,Ferretti:2018ojb,Zeminiani:2021xvw,Belyaev:2006vn,Belyaev:2007yc,Belyaev:2009ag,Sofianos:2010,Yokota:2013sfa}.

However, no experimental evidence has been found for a ${\phi N}$ bound state in photon-production measurements. 
The LEPS collaboration measured the differential cross section of photoproduction of $\phi$ mesons on protons~\cite{LEPS:2005hax}. Based on these data, Titov \etal~\cite{Titov:2007xb} utilized vector meson dominance (VMD) assumption and obtained $a_{\phi p}\approx-0.15\ \mathrm{fm}$, which is consistent with the scattering length derived with the QCD sum rule analysis \cite{Koike:1996ga}. The CLAS collaboration measured the near-threshold differential cross section for the $\gamma p \rightarrow \phi p$ process, covering center-of-mass energies from 1.97 to 2.84 GeV~\cite{Dey:2014tfa}. Strakovsky \etal applied the VMD approach to the CLAS measurements, yielding $a_{\phi p} \approx -0.063 \pm 0.010\ \mathrm{fm}$~\cite{Strakovsky:2020uqs}.

Until 2021, the ALICE collaboration analyzed the $p\phi$ correlation function in proton-proton collisions, revealing for the first time a strong attraction between $\phi$ and nucleons, with a real part of the scattering length $Re[a_{\phi p}]= -0.85 \pm 0.34 (stat.) \pm 0.14 (syst.) \ \mathrm{fm}$~\cite{ALICE:2021cpv}. Furthermore, ALICE proposed that elastic contributions dominate in $\phi N$ interactions in the vacuum, which contrasts with earlier measurements of $\phi$ absorption off various nuclear targets in different reactions~\cite{CLAS:2010pxs,Polyanskiy:2010tj,HADES:2018qkj}.

For the charmonium sector, a scattering length $a_{J/\psi p}=-(3.08 \pm 0.55(stat.)\pm 0.45(syst.)) \times 10^{-3}\ \mathrm{fm}$ is extracted from the near-threshold photon-production total cross section using GlueX data~\cite{GlueX:2019mkq} as reported by Strakovsky \etal~\cite{Strakovsky:2019bev}. Another value of $a_{J/\psi p}=-0.046 \pm 0.005\ \mathrm{fm}$ is derived from a global fit to both differential and total cross sections of earlier photon-production data~\cite{Gryniuk:2016mpk}. The above two results in Refs.~\cite{Gryniuk:2016mpk,Strakovsky:2019bev} were both extracted within the VMD framework. Recently, JPAC combined the latest GlueX and $J/\psi-007$ measurements of near-threshold $J/\psi$ photoproduction cross sections with constraints from the low-energy unitarity, yielding a scattering length $a_{J/\psi p} \approx O(1)\ \mathrm{fm}$~\cite{JointPhysicsAnalysisCenter:2023qgg}, where the VMD was not adopted.

In summary, no experimental evidence has been found for the existence of $\phi p$ or $J/\psi p$ bound states. Moreover, significant inconsistencies persist among the reported scattering lengths. The discrepancies in $\phi N$ scattering lengths between photon production processes and proton-proton collisions may arise from differences in the measurement methods. Alternatively, the smaller scattering lengths observed in the photoproduction of $\phi p$~\cite{Titov:2007xb,Strakovsky:2020uqs} and $J/\psi p$~\cite{Gryniuk:2016mpk,Strakovsky:2019bev} within the VMD framework, compared to those obtained without the VMD assumption~\cite{ALICE:2021cpv,JointPhysicsAnalysisCenter:2023qgg}, could be attributed to the off-shell effects of the vector meson~\cite{Pentchev:2020kao} inherent in the VMD framework. Another limitation is that all the above experimental measurements are restricted to spin-averaged results.  Therefore, novel inputs are essential for a more comprehensive understanding of these issues.

Recently, the Lattice QCD started from the first principle and made progresses on the $J/\psi N$, $\eta_cN$ and $\phi N$ interactions. In 2022, the HAL QCD collaboration extracted the interaction potential for $\phi N$ in the ${}^4S_{3/2}$ channel for the first time near the physical pion mass~\cite{Lyu:2022imf}. Their results showed that the $\phi N$(${}^4S_{3/2}$) interaction is attractive at all ranges, with scattering length $a_{\phi p}=-1.43(23)_{\mathrm{stat}}\binom{+36}{-06}_{\mathrm{syst}}\ \mathrm{fm}$. Recently, HAL QCD also presented their new results on nucleon-charmonium interactions near the physical pion mass~\cite{Lyu:2024ttm}. They found attraction in both the ${}^4S_{3/2}$ and ${}^2S_{1/2}$ nucleon-charmonium channels, with the scattering lengths given by $a_{J/\psi N}^{{}^4S_{3/2}}=-0.30(2)\binom{+2}{-0}\ \mathrm{fm}$ and $a_{J/\psi N}^{{}^2S_{1/2}}=-0.38(4)\binom{+3}{-0}\ \mathrm{fm}$, which are consistent with some theoretical calculations~\cite{Hayashigaki:1998ey,Sibirtsev:2005ex,Wu:2024xwy} but are explicitly larger than those obtained from photoproduction experiments.
In addition, several previous lattice QCD studies have investigated charmonium-nucleon interactions using unphysical pion masses~\cite{Yokokawa:2006td,Kawanai:2010ru,Skerbis:2018lew,Sugiura:2019pye}.

Due to its strong coupling with the $\Lambda K$ and $\Sigma K$ channels, HAL QCD is unable to effectively extract the interaction of the $\phi N$(${}^2S_{1/2}$) channel. To address this issue, one approach involves using the potential to fit the correlation function~\cite{Chizzali:2022pjd}, which claims the existence of a deep bound state in the $\phi N$(${}^2S_{1/2}$) channel. In this work, we refer this interaction as Model A.  However, it is important to note that the same correlation function can also be fitted well by coupled channel dynamics~\cite{Feijoo:2024bvn,Abreu:2024qqo}, without requiring a bound state predominantly composed of $\phi N $. To investigate other possibilities, in this work, we propose a new potential for the $\phi N({}^2S_{1/2})$ interaction that connects it with the interactions of $\phi N({}^4S_{3/2})$, $J/\psi N$(${}^4S_{3/2}$) and $J/\psi N$(${}^2S_{1/2}$), referred as Model B.

In addition to two-body systems, few-body systems also hold significant implications for understanding hadronic interactions, as reviewed in Ref.~\cite{Liu:2024uxn}. For instance, the existence of a $DDD^*$ bound state would impose strong constraints on the $DD^*$ and $D\bar{D}^*$ interactions, providing insights into the pole position of the $Z_c(3900)$~\cite{Zhu:2024hgm}. In the $\phi NN$, $J/\psi NN$, and $\eta_c NN$ three-body systems, several pioneering studies have been conducted. Notably, Belyaev et al. used Faddeev equations in differential form \cite{Belyaev:2007yc,Belyaev:2009ag}, while Sofianos et al. \cite{Sofianos:2010} employed a two-variable integro-differential equation based on the Faddeev expansion of the wavefunction to conduct advanced research on $\phi NN$. Based on the $\phi N$ interaction proposed by Gao \etal~\cite{Gao:2000az}, both Belyaev et al. and Sofianos \etal found $\phi NN$ bound states. Yokota et al. applied the Gaussian Expansion Method (GEM) to study $J/\psi NN$ and $\eta_c NN$~\cite{Yokota:2013sfa}, concluding that the small scattering length obtained from quenched Lattice QCD~\cite{Kawanai:2010ru} cannot support a three-body charmonium nucleus bound state. Additionally, Belyaev \etal used the AGS equation to investigate $\eta_c NN$~\cite{Belyaev:2006vn}, predicting a shallow bound state based on the potential proposed by Brodsky \etal \cite{Brodsky:1989jd}. Etminan and Aalimi solved the Faddeev equation with hyperspherical harmonics~\cite{Etminan:2024vkv}.
However, the interactions between heavy quarkonium and nucleons used in these studies are relatively crude, precluding definitive conclusions based on these potentials. 

The HAL QCD potential for $\phi N$ near the physical pion mass  revealed that neither the Yukawa-type nor power-type potentials used in previous studies accurately describe the hadronic interaction. Instead, the two-pion exchange potential dominates at long range. Although none of the extracted potentials identified any two-body bound states in $\phi N$, $J/\psi N$, or $\eta_c N$ systems, the potential with a scattering length smaller than -1 fm has attracted attention for investigating three-body bound states in $\phi NN$ systems. Recently,  Filikhin \etal conducted a comprehensive analysis of $I(J^P)=0(0^-)$, $0(1^-)$, $1(1^-)$, and $0(2^-)$ $\phi NN$ systems using Faddeev equations in configuration space~\cite{Filikhin:2024avj}. They identified bound states in the $0(0^-)$, $0(1^-)$, and $1(1^-)$ systems using the HAL QCD and Model A potentials. A key issue is that Filikhin \etal used a single-channel spin-averaged potential for the $0(1^-)$ and $1(1^-)$ $\phi NN$ channels. 
However, in Model A, the interactions in the $\phi N({}^4S_{3/2})$ and $\phi N({}^2S_{1/2})$ channels differ significantly. It is more reasonable to account for the complex coupling effects between spin channels. Additionally, the $\phi N({}^2S_{1/2})$ potential, determined using correlation functions, neglects the coupled-channel effects, making its uncertainty uncontrollable. Exploring alternative models for the $\phi N({}^2S_{1/2})$ interaction, as well as investigating the $\phi NN$ system within new interactions, would be of great significance. Meanwhile, it is also crucial to investigate the bound states and resonances of the $J/\psi NN$, $\eta_c NN$ and charmonium-nucleus systems.

% using the $\phi N({}^4S_{3/2})$ potential developed by HAL QCD~\cite{Lyu:2022imf}, Etminan and Aalimi solved the Faddeev equation with hyperspherical harmonics and found a bound state of $I(J^P)=0(2^-)$ with a binding energy of approximately 7 MeV~\cite{Etminan:2024vkv}. Shortly thereafter, but did not find a bound state in the $0(2^-)$ $\phi NN$ system, which contradicts the findings of Etminan and Aalimi. 

% The few-body systems hold significant implications for the study of exotic hadrons and the spectrum of hadronic states. Recent experimental discoveries, such as the identification of pentaquarks and tetraquarks, highlight the necessity for a deeper understanding of multi-quark systems~\cite{Chen:2016qju,Liu:2019zoy,Chen:2022asf}. These exotic states challenge the conventional quark model and motivate theoretical frameworks to accurately describe their properties and interactions~\cite{Meng:2021jnw,Wang:2019ato,Wang:2019nvm,Meng:2022ozq,Lin:2024qcq,Guo:2017jvc,Wu:2024zbx,Ma:2024vsi,Liu:2009qhy,Liu:2008xz}. Furthermore, three-body systems possess certain unique properties, such as the Efimov effect and three-body forces, which do not occur in two-body systems. These characteristics merit further investigation~\cite{Kalantar-Nayestanaki:2011rzs,Endo:2024cbz}.

In this work we perform complete dynamical calculations on the  $\phi NN$, $J/\psi NN$ and $\eta_c NN$ systems using the complex scaling method (CSM)~\cite{Aguilar:1971ve,Balslev:1971vb,Moiseyev:1998gjp,Aoyama:2006hrz,Carbonell:2013ywa,Hiyama:2016nwn,Dote:2017wkk} and Gaussian expansion method (GEM)~\cite{Hiyama:2003cu}. Beyond the bound state, the CSM enables the study of resonances if they exist in these systems. The theoretical frameworks accurately describe the properties of multi-quark systems, few-hadron systems and few-lepton systems~\cite{Lin:2022wmj,Happ:2023kcc,Meng:2023jqk,Chen:2023eri,Chen:2023syh,Meng:2021jnw,Meng:2024yhu,Wu:2024euj,Wu:2024zbx,Lin:2024qcq,Wu:2024hrv,Zhu:2024hgm,Wu:2024ocq,Ma:2024vsi,Ma:2025rvj,Yang:2025wqo}.

In this study, we employ the HAL QCD potential to investigate the $\phi NN$ system with the highest spin of $0(2^-)$, the $J/\psi NN$ system with $0(0^-)$, $0(1^-)$, $1(1^-)$, and $0(2^-)$, and the $\eta_c NN$ system with $0(1^-)$ and $1(0^-)$. For the $S=1$ and $S=0$ $\phi NN$ channels, we utilize the two aforementioned models for comparison. Furthermore, we provide an approximate estimation of the minimal mass number of a nucleus required for the formation of bound charmonium-nucleus states.

This paper is organized as follows:  
In Sec.~\ref{sec:framework}, we outline the theoretical framework, including the Hamiltonian, wave function construction, complex scaling method, and the approach for analyzing spatial structures.  In Sec.~\ref{sec:2body}, we explore the possible two-body bound states and the scattering lengths and effective ranges.
In Sec.~\ref{sec:results_phiNN}, we present the numerical results for the $\phi NN$ system.  
In Sec.~\ref{sec:charmonium}, we analyze the results for the $J/\psi NN$ and $\eta_c NN$ systems.  
Finally, in Sec.~\ref{sec:sum}, we provide a brief summary and discussion.  

\section{Theoretical framework}~\label{sec:framework}

\subsection{Hamiltonian}~\label{subsec:Hamiltonian}

The nonrelativistic Hamiltonian of a three-body hadron system can be expressed as
\begin{align}\label{eq:Hamiltonian}
H=\sum_i^3\left(m_i+\frac{\boldsymbol{p}_i^2}{2m_i}\right)+\sum_{i<j=1}^3 V_{i j}\,,
\end{align}
where $m_i$ and $\boldsymbol{p}_i$ are the mass and momentum of hadron $i$. In Sec.~\ref{sec:results_phiNN}, the typical binding energy of bound states is $2\sim50\ \mathrm{MeV}$, corresponding to a characteristic momentum of $40\sim150\ \mathrm{MeV}$. For $\phi NN$ systems, the mass scale is 1 GeV, giving $p/m \approx 0.1$. In charmonium-nucleus systems, $p/m$ is much smaller than that in $\phi NN$. Thus, non-relativistic kinematics in the Hamiltonian remain appropriate.

For nucleon nucleon interactions, we utilize the modified  Malfliet-Tjon (MT) potential~\cite{Malfliet:1968tj,Friar:1990zza}. It only incorporates S-wave Yukawa-type potentials characterized by a short-range repulsion and a long-range attraction. The $N N$ potential reads 
\begin{align}\label{eq:NN}
   V^{{}^1S_{0}}_{N N}(r)=\sum_{i=1}^2 C^{{}^1S_{0}}_i \frac{e^{-\mu_i r}}{r} , \\
   V^{{}^3S_{1}}_{N N}(r)=\sum_{i=1}^2 C^{{}^3S_{1}}_i \frac{e^{-\mu_i r}}{r}.
\end{align}
The parameters are given in Table~\ref{tab:paraNN}.

For $J/\psi N$ and $\eta_cN$ interactions, we adopt the potential from HAL QCD in Ref.~\cite{Lyu:2024ttm}, which employed three Gaussian functions for fitting. The potentials are expressed as follows,
\begin{align}\label{eq:jpsietacN}
V^{{}^4S_{3/2}}_{J/\psi N}(r)=& \sum_{i=1}^3 a_{i1} e^{-\left(r / b_{i1}\right)^2} ,    \\
V^{{}^2S_{1/2}}_{J/\psi N}(r)=& \sum_{i=1}^3 a_{i2} e^{-\left(r / b_{i2}\right)^2} ,    \\
V^{{}^2S_{1/2}}_{\eta_c N}(r)=& \sum_{i=1}^3 a_{i3} e^{-\left(r / b_{i3}\right)^2} .  
\end{align}
The parameters are collected in Table~\ref{tab:parajpsiN}.

For the $\phi N$ spin quartet interaction, we utilize the HAL QCD potential~\cite{Lyu:2022imf}, which incorporates a combination of two Gaussian potentials and the two-pion exchange potential. The potential reads,
\begin{align}\label{eq:phiN3}
V^{{}^4S_{3/2}}_{\phi N}(r)= \sum_{i=1}^2 \alpha_i e^{-\left(r / \beta_i\right)^2}&+\alpha_3 m_\pi^4 f\left(r ; \beta_3\right)\left(\frac{e^{-m_\pi r}}{r}\right)^2 ,   \\
f\left(r ; \beta_3\right)=&\left(1-e^{-\left(r / \beta_3\right)^2}\right)^2.
\end{align}
The parameters are shown in Table~\ref{tab:paraphiN}.
% \begin{align}\label{AP1}
%     V_{i j}=-\frac{3}{16}&\lambda_i^c \cdot \lambda_j^c\Big(-\frac{\kappa}{r_{ij}}+\lambda r_{i j}^{2 / 3}-\Lambda\nonumber\\
%     &+\frac{8\pi\kappa'}{3m_{i}m_{j}}\frac{\exp(-r_{ij}^{2}/r_{0}^{2})}{\pi^{3/2}r_{0}^{3}}\boldsymbol{s}_{i}\cdot\boldsymbol{s}_{j}
%     \Big),
% \end{align}

For the spin doublet potential, no direct lattice QCD results are available due to the complexity of the coupled channel effect. As noted in Sec.~\ref{sec:intro}, the potential extracted from the combined analysis of the correlation function and lattice QCD spin-$3/2$ data also neglects the coupled channel effect, which may compromise the reliability of the potential.  Therefore, in spin-$1/2$ systems, we explore two different models to investigate alternative possibilities.
In Model A, a spin-$1/2$ potential function is constructed by modifying Eq.~\eqref{eq:phiN3} with an additional factor applied to the Gaussian part~\cite{Chizzali:2022pjd}. \begin{align}\label{eq:phiN1A}
V^{{}^2S_{1/2},A}_{\phi N}(r)=&\beta \sum_{i=1}^2 \alpha_i e^{-\left(r / \beta_i\right)^2}
+\alpha_3 m_\pi^4 f\left(r ; \beta_3\right)\left(\frac{e^{-m_\pi r}}{r}\right)^2,        
\end{align}
The parameter $\beta$ is determined by fitting the $p-\phi$ correlation functions, while other parameters adopt the values from the HAL QCD spin-$3/2$ potential~\cite{Chizzali:2022pjd,Lyu:2022imf}. The value of $\beta$ is shown in Table~\ref{tab:paraphiN}. It is worth noting that the lattice QCD calculations are not performed at the physical pion mass but rather at a value close to it. The Model A is constructed by directly varying the explicit pion mass in Eq.~\eqref{eq:phiN1A} without accounting for the pion-mass dependence of the parameters.

In Model B, we construct the spin-$1\over 2$ $\phi N$ potential by relating it to that of the $J/\psi N$ system without using $p$-$\phi$ correlation function data. We assume that the spin-spin part in the $J/\psi N$ and $\phi N$ interaction is inversely proportional to the hadron mass. Using this assumption, we can derive the $\phi N$ doublet potential as follows,
\begin{align}
V^{Spin}_{J/\psi N}(r)=&\frac{V^{{}^4S_{3/2}}_{J/\psi N}-V^{{}^2S_{1/2}}_{J/\psi N}}{3}  , \label{eq:phiNBspin}\\
V^{Spin,B}_{\phi N}(r)=&\frac{m_{J/\psi}}{m_\phi}V^{Spin}_{J/\psi N}(r),\\
V^{{}^2S_{1/2},B}_{\phi N}(r)=&V^{{}^4S_{3/2}}_{\phi N}(r)-3 V^{Spin,B}_{\phi N}(r) .\label{eq:phiN1B}
\end{align}
where the spin-spin interaction is labeled by the superscript ``Spin".

\begin{table}[htbp]
    \centering
    \caption{The parameters of the modified Malfliet-Tjon (MT) model.}
    \label{tab:paraNN}
    \begin{tabular*}{\hsize}{@{}@{\extracolsep{\fill}}cccccc@{}}
\hline\hline  & $C_1[\mathrm{MeV} \cdot \mathrm{fm}]$ & $\mu_1[\mathrm{fm}^{-1}]$ & $C_2[\mathrm{MeV} \cdot \mathrm{fm}]$ & $\mu_2[\mathrm{fm}^{-1}]$ \\
\hline ${}^1S_0$  & -514 & 1.55 & 1439 & 3.11 \\ 
\hline ${}^3S_1$  & -627 & 1.55 & 1439 & 3.11 \\
\hline\hline
    \end{tabular*}
\end{table}

\begin{table}[htbp]
    \centering
    \caption{The parameters of ${c\bar{c}}N$ potential $a_1,a_2,a_3$ (in $\mathrm{MeV}$), and $b_1,b_2,b_3 $ (in $\mathrm{fm}$) are taken from Ref.~\cite{Lyu:2024ttm}.}
    \label{tab:parajpsiN}
    \begin{tabular*}{\hsize}{@{}@{\extracolsep{\fill}}cccccccccc@{}}
\hline\hline
&  & $a_1$ & $b_1$ 
& $a_2$ & $b_2$  & $a_3$ & $b_3$  \\
\hline & $J/\psi N({}^4S_{3/2})$ & -51(1) & 0.09(1) & -13(6) & 0.49(7) & -22(5) & 0.82(6)  \\
\hline & $J/\psi N({}^2S_{1/2})$ & -101(1) & 0.13(1) & -33(6) & 0.44(5) & -23(8) & 0.83(9)  \\
\hline & $\eta_c N({}^2S_{1/2})$ & -264(14) & 0.11(1) & -28(13) & 0.24(6) & -22(2) & 0.77(3)  \\
\hline\hline
    \end{tabular*}
\end{table}

\begin{table}[htbp]
    \centering
    \caption{The parameters of $\phi N$ potential $\alpha_1,\alpha_2$ (in $\mathrm{MeV}$), and $\beta_1,\beta_2,\beta_3$, as well as $\alpha_3m_\pi^4 $ (in $\mathrm{fm}$) are taken from Refs.~\cite{Lyu:2022imf,Chizzali:2022pjd}.}
    \label{tab:paraphiN}
    \begin{tabular*}{\hsize}{@{}@{\extracolsep{\fill}}ccccccccccc@{}}
\hline\hline
&  $\alpha_1$ & $\beta_1$ 
& $\alpha_2$ &$\beta_2$  & $\alpha_3 m_\pi^{4 }$ & $\beta_3$ &$\beta$  \\
\hline  & -371(27) & 0.13(1) & -119(39) & 0.30(5) & -97(14) & 0.63(4) & 6.9\\
\hline\hline
    \end{tabular*}
\end{table}

Most of the above potentials are derived from HAL QCD results. Since the lattice QCD simulations are performed at an unphysical pion mass, our calculations will also use the corresponding unphysical hadron masses, as listed in Table~\ref{tab:mass}. In the charm sector, lattice QCD employs two sets of parameters to interpolate the physical mass of the charm quark and reproduce the dispersion relation for the spin-averaged 1S charmonium state~\cite{Lyu:2024ttm}. Consequently, we adopt the experimental masses of $J/\psi$ and $\eta_c$. It is worth emphasizing that these masses closely approximate their physical counterparts.

\begin{table}[htbp]
    \centering
    \caption{The masses of $\phi,N,\pi$ (in $\mathrm{GeV}$) used in HAL QCD are taken from Refs.~\cite{Lyu:2022imf,Lyu:2024ttm}, while the masses of $J/\psi,\eta_c$ (in $\mathrm{GeV}$) are taken from experiments.}
    \label{tab:mass}
    \begin{tabular*}{\hsize}{@{}@{\extracolsep{\fill}}ccccccccccc@{}}
\hline\hline
&  $m_N$ & $m_\pi$ & $m_\phi$
& $m_{J/\psi}$ & $m_{\eta_c}$ \\
\hline & 0.954 & 0.146 & 1.048 & 3.096 & 2.984 \\
\hline\hline
    \end{tabular*}
\end{table}

\subsection{Wave function construction}~\label{subsec:wavefunction}

The single hadron wave function space is the direct product of spatial space $\chi_{r} $, spin space $\chi_{s}$, and isospin space $\chi_{f}$.
\begin{equation}\label{eq:Abasis}
\psi=\mathcal{A}\left(\chi_{r} \otimes \chi_{s} \otimes \chi_{f}\right),
\end{equation}
where $\mathcal{A}$ is the anti-symmetrization operator, representing the exchange of two nucleons. For the $\phi NN$ system, $\mathcal{A}=\left(1-P_{23}\right)$, where $P_{i j}$ permutes the $i$th and $j$th hadron. 

For the spatial wave function, the Gaussian expansion method (GEM)~\cite{Hiyama:2003cu} is employed. Namely, the spatial wave function is expanded using the following basis:
\begin{equation}\label{eq:basisSpace}
\chi_{n l m}(\boldsymbol{r})=\sqrt{\frac{2^{l+5 / 2}}{\Gamma\left(l+\frac{3}{2}\right) r_n^3}}\left(\frac{r}{r_n}\right)^l e^{-\frac{r^2}{r_n^2}} Y_{l m}(\hat{r}),
\end{equation}
where the $r_n$ is taken in geometric progression, $r_n=r_0 a^{n-1}$. $Y_{l m}$ is the spherical harmonics.

\begin{figure}[htbp]
  \centering
  \includegraphics[width=0.47\textwidth]{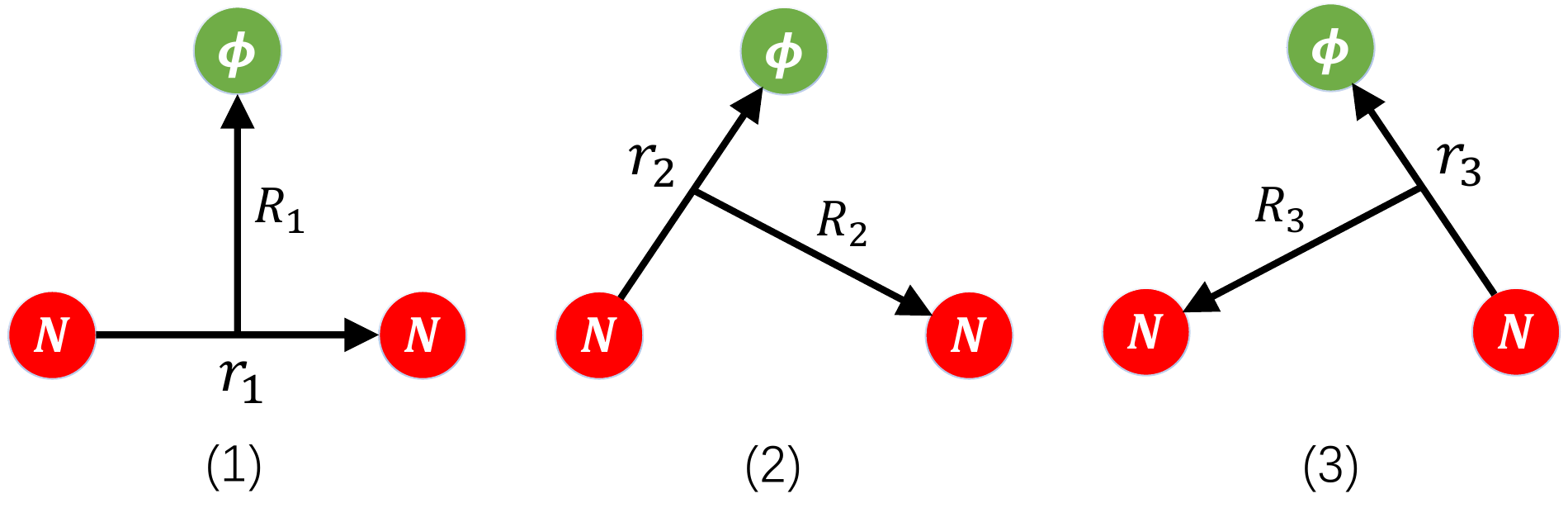} 
  \caption{\label{fig:structure} The Jacobi coordinate configurations of the $\phi NN$ system. }
    \setlength{\belowdisplayskip}{1pt}
\end{figure}

The total orbital angular momentum serves as a good quantum number due to the spatial rotational invariance of each two-body potential. We focus on $L=0$ states as they correspond to the low energy levels. Since the interactions we employ are purely S-wave, the S-wave Gaussian bases corresponding to the three distinct Jacobi configurations illustrated in Fig.~\ref{fig:structure} form a complete set for expanding the spatial wave function. We employ subscripts 1, 2, and 3 to distinguish the three distinct Jacobi coordinate configurations.

In our calculations, we utilize 20 basis functions per coordinate, with $n_{max}=20$. The configuration of the basis parameters is as follows,
\begin{equation}
r_0=0.02\ \mathrm{fm}, r_{n_{{max}}}=40.0\ \mathrm{fm},
\end{equation}
where we use the same parameters of Gaussian basis $r_0,r_{n_{{max}}}$ for different Jacobi coordinates.

\clabel[S-wave projection]{It is worth noting that the GEM was often employed for the local potential, contributing to all partial waves, as illustrated in refs.~\cite{Meng:2023jqk,Chen:2023eri,Chen:2023syh,Meng:2021jnw,Meng:2024yhu,Wu:2024euj,Wu:2024zbx,Lin:2024qcq,Wu:2024hrv,Zhu:2024hgm,Wu:2024ocq,Ma:2024vsi,Ma:2025rvj,Yang:2025wqo}. However, in this work, the interaction described in Sec.~~\ref{sec:2body} is restricted to the S-wave only. We lack sufficient experimental data to constrain the potential for higher partial waves. Since our focus is exclusively on the ground state, the influence of higher partial waves is neglected. Utilizing the S-wave-projected interaction, however, introduces some computational complexities when applied to the GEM framework. For any given potential $\hat{V}$, it can be expanded in the angular momentum basis $|r L M\rangle$ as follows, 
\begin{equation}
\hat{V}=\sum_{L,M,L',M'}\int r^2dr r'^2dr'|rLM\rangle \langle rLM| \hat{V} |r'L'M'\rangle  \langle r'L'M'|,
\end{equation}
When employing an S-wave potential, we obtain,
\begin{equation}\label{S-wave potential}
    \begin{aligned}
    \langle rLM| \hat{V} |r'L'M'\rangle&=\delta_{L,0}\delta_{L',0} V(r)\frac{\delta(r-r')}{r^2},  \\
\hat{V}&=\int r^2dr |r00\rangle V(r)   \langle r00|
    \end{aligned}
\end{equation}
When working with the Jacobi configurations of $|\chi_3\rangle$, which are identical to the configuration of the potential $V_3(r_3)$, the form of the matrix element of the S-wave potential is the same as that of a typical local potential,
\begin{equation}\label{eq:same configuration}
\begin{aligned}
\left\langle\chi_3\right| \hat{V_3}\left|\chi_2\right\rangle  
=&\int r_3^2 dr_3\left\langle\chi_3\right| r_300\rangle \langle r_300 \left|\chi_2\right\rangle V(r_3)\\
=&\sum_{L,M}\int r_3^2 d r_3\left\langle\chi_3 \mid r_3 LM\right\rangle V_{3}(r_3) \left\langle r_3 LM \mid \chi_2\right\rangle \\
=&\int d \vec{r_3}\left\langle\chi_3 \mid \vec{r_3}\right\rangle V_{3}(r_3) \left\langle \vec{r_3} \mid \chi_2\right\rangle.
\end{aligned}
\end{equation}
The second equality arises because the wave function $\langle\chi_3|r_3LM\rangle$ is purely S-wave, and only $\langle\chi_3|r_300\rangle$ survives.  
When working with the Jacobi configurations of $\left|\chi_1\right\rangle$ and $\left|\chi_2\right\rangle$, which differ from the configuration of the potential $V_3\left(r_3\right)$, it is necessary to transform the Gaussian basis functions to align with the potential's configuration $\left|r_3 LM\right\rangle$ to calculate the corresponding matrix elements. However, transforming an S-wave basis from its original configuration to a new one via Jacobian transformations introduces higher partial wave components. Given that the extracted potential is purely S-wave, we must project the wave function onto the S-wave configuration of the potential $\left| r_3 0 0 \right\rangle$, as follows,
\begin{equation}\label{eq:wavefunction projection}
\begin{aligned}
\langle \chi_1|r_300\rangle&=\int d\vec{r'_1}d\vec{r'_3}\langle \chi_1| \vec{r'_1}\rangle\langle \vec{r'_1}|\vec{r'_3}\rangle\langle \vec{r'_3}|r_300\rangle\\
&=\int d\hat{r_3}\langle \chi_1|\vec{r_3}\rangle Y_{00}(\hat{r_3})
\end{aligned}
\end{equation}
The second step involves obtaining $\langle \chi_1|\vec{r'_3}\rangle$ through a coordinate transformation, while the final step integrates out the angular dependence of the newly transformed wave function to achieve the S-wave projection. Finally, the matrix element can be expressed as,  
\begin{equation}\label{eq:different configuration}
\begin{aligned}
\left\langle\chi_1\right| \hat{V_3}\left|\chi_2\right\rangle  
=\int r_3^2 dr_3\left\langle\chi_1\right| r_300\rangle \langle r_300 \left|\chi_2\right\rangle V_3(r_3)\\
=\int r_3^2 d r_3\left\langle\chi_1 \mid r_3 00\right\rangle V_{3}(r_3) \left\langle r_3 00 \mid \chi_2\right\rangle.
\end{aligned}
\end{equation}}

For the spin-flavor wave functions $[(NN)_{S_{NN},I_{NN}}\phi]_{S,I}$ and $[(\phi N)_{S_{\phi N},I_{\phi N}}N]_{S,I}$, we have included all possible basis functions that align with the corresponding S-wave Jacobi configurations. The inner products of these discrete wave functions for total spin $S=0,1,2$ and total isospin $I=0,1$ are detailed in Appendix~\ref{AppendixB}.

\subsection{Complex scaling method}~\label{subsec:method}

Few-body systems encompass sub-channels such as break-up and rearrangement channels. Near-threshold bound states or resonant states can be determined using the complex scaling method (CSM). The CSM provides a direct approach for calculating the energies and decay widths of resonant states in many-body systems by performing an analytical continuation of the Schr\"odinger equation~\cite{Aguilar:1971ve,Balslev:1971vb,aoyama2006complex}. This is accomplished by applying a complex rotation to the coordinate $\boldsymbol{r}$ and momentum $\boldsymbol{p}$, given by
\begin{align}\label{eq:complexRotation}
U(\theta) \boldsymbol{r}=\boldsymbol{r} e^{i \theta}, \quad U(\theta) \boldsymbol{p}=\boldsymbol{p} e^{-i \theta}.
\end{align}
Under the rotation, the Hamiltonian in Eq. (\ref{eq:Hamiltonian}) becomes
\begin{equation}\label{eq:HamiltonianComplex}
H(\theta)=\sum_{i=1}^3\left(m_i+\frac{p_i^2 e^{-2 i \theta}}{2 m_i}\right)+\sum_{i<j=1}^3 V_{i j}\left(r_{i j} e^{i \theta}\right).
\end{equation}
Meanwhile, for the resonant states with pole positions within the range of the rotated angle, their wave functions become normalizable by integration, thereby solvable through localized Gaussian bases in the same way as bound states. As a result, solving the complex-scaled Schr\"odinger equation will simultaneously yield the eigenenergies of bound states and resonant states within the rotated angle.

A typical pattern of the solved eigenenergies in the complex energy plane is: The bound states lie on the negative real axis of the energy plane. The continuum states align along beams originating from thresholds with $\operatorname{Arg}(E)=-2 \theta$. The resonant states with mass $M_R$ and width $\Gamma_R$ are located at $E_R=M_R-i \Gamma_R / 2$, and only those within $\left|\operatorname{Arg}\left(E_R\right)\right|<2 \theta$ can be solved. The positions of the bound and resonant states remain unchanged with the variation of the rotation angle. The CSM is therefore an effective tool for distinguishing between scattering states and near-threshold bound states. One can find more details in Refs.~\cite{Lin:2022wmj,Chen:2023eri,Chen:2023syh}.
\subsection{Spatial structure}

The root-mean-square (rms) radius is a good physical quantity for reflecting the spatial structure of the hadron states. For near-threshold states, this is also an auxiliary criterion for determining whether it is a bound state or a scattering state.

The definition of the rms radius under CSM is 
\begin{equation}
r_{i j}^{{rms,C}} \equiv \operatorname{Re}\left[\sqrt{\frac{\left(\Psi(\theta)\left|r_{i j}^2 e^{2 i \theta}\right| \Psi(\theta)\right)}{\left(\Psi(\theta) \mid \Psi(\theta)\right)}}\right],
\end{equation}
where the $\Psi(\theta)$ is the obtained complex wave function of the state. The round bra-ket represents the so-called c-product~\cite{Romo:1968tcz} defined as
\begin{equation}
\left(\phi_n \mid \phi_m\right) \equiv \int \phi_n(\boldsymbol{r}) \phi_m(\boldsymbol{r}) {d}^3\boldsymbol{r},
\end{equation}
without taking complex conjugate of the bra-state. This procedure ensures the function inside the integral is analytic, thereby the expectation value of the physical quantity remains stable as the rotation angle changes. The rms radius calculated from the c-product is generally not real for resonance. However, it is real for bound state, as discussed in Ref.~\cite{homma1997matrix}.

\change{The one-body density distribution is also a physical quantity that reflects the radial distributions of hadrons, where $\hat{r_i}$ represents the radius operator of the i-th particle relative to the center of mass~\cite{MLP}.
\begin{equation}
    \rho(r)=\langle \Psi| \Sigma_i \ \delta(r-\hat{r}_i) \ |\Psi \rangle
\end{equation}}

\section{Two-body systems}~\label{sec:2body}

For each two-body channel, we solve the Schr\"odinger equation to determine the bound state, including its theoretical binding energy and root-mean-square (rms) radii. The scattering parameters are obtained by solving the Lippmann-Schwinger equation. The corresponding values are provided in Table~\ref{tab:twobody}.

\begin{table}[htbp]
\renewcommand{\arraystretch}{1.4}
\centering
\caption{\label{tab:twobody} The bound states (BS), including their binding energies, rms radii, scattering lengths, and effective ranges for two-body systems, are presented below. }
\begin{tabular*}{\hsize}{@{}@{\extracolsep{\fill}}lcccccc@{}}

\hline\hline
  &BS & $E_{{Theo.}}[\mathrm{MeV}]$  & $r^{{rms}}_{{Theo.}}[\mathrm{fm}]$ & $a_0[\mathrm{fm}]$ & $r_{eff}[\mathrm{fm}]$ \\
 \hline 
 ${NN({}^1S_0)}$ & - & - & - & -34.14 & 3.09\\
 ${NN({}^3S_1)}$ & $d$ & $2.41$ & $3.84$ & 5.28 & 1.94\\
 ${\phi N ({}^4S_{3/2})}$  & - & - & - & -1.44 & 2.36\\
 ${\phi N ({}^2S_{1/2}),A}\ \ $ & \change{$\tilde{N}$} & 18.4  & 1.27 & 1.76 & 0.45\\
 ${\phi N ({}^2S_{1/2}),B}\ \ $ & - & -  & - & -2.19 & 1.88\\
 ${J/\psi N ({}^4S_{3/2})}$  & - & - & - & -0.30 & 3.28\\
 ${J/\psi N ({}^2S_{1/2})}$  & - & - & - & -0.38 & 2.67\\
 ${\eta_c N ({}^2S_{1/2})}$  & - & - & - & -0.22 & 3.61\\
\hline \hline
\end{tabular*}
\end{table}

\begin{figure*}[]
\centering
  \includegraphics[width=0.83\textwidth]{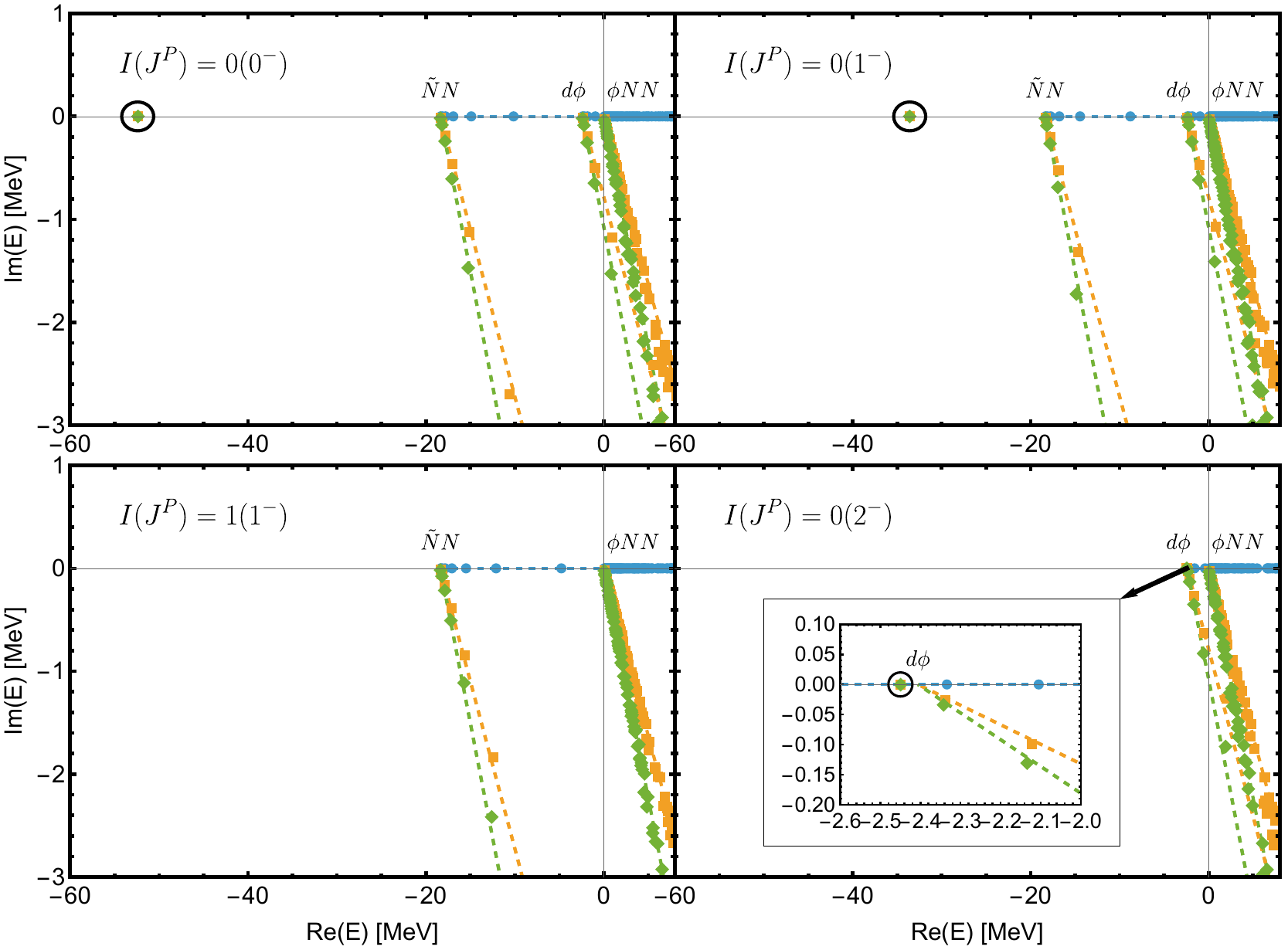} 
  \caption{\label{fig:phiNNA}The complex energy eigenvalues of the $\phi NN$ states in Model A with varying $\theta$ in the CSM. The dashed lines represent the continuum lines rotating along ${Arg}(E)=-2\theta$. The bound states do not shift as $\theta$ changes and are marked out by black circles.}
    \setlength{\belowdisplayskip}{1pt}
\end{figure*}

\begin{figure*}[]
\centering
  \includegraphics[width=0.83\textwidth]{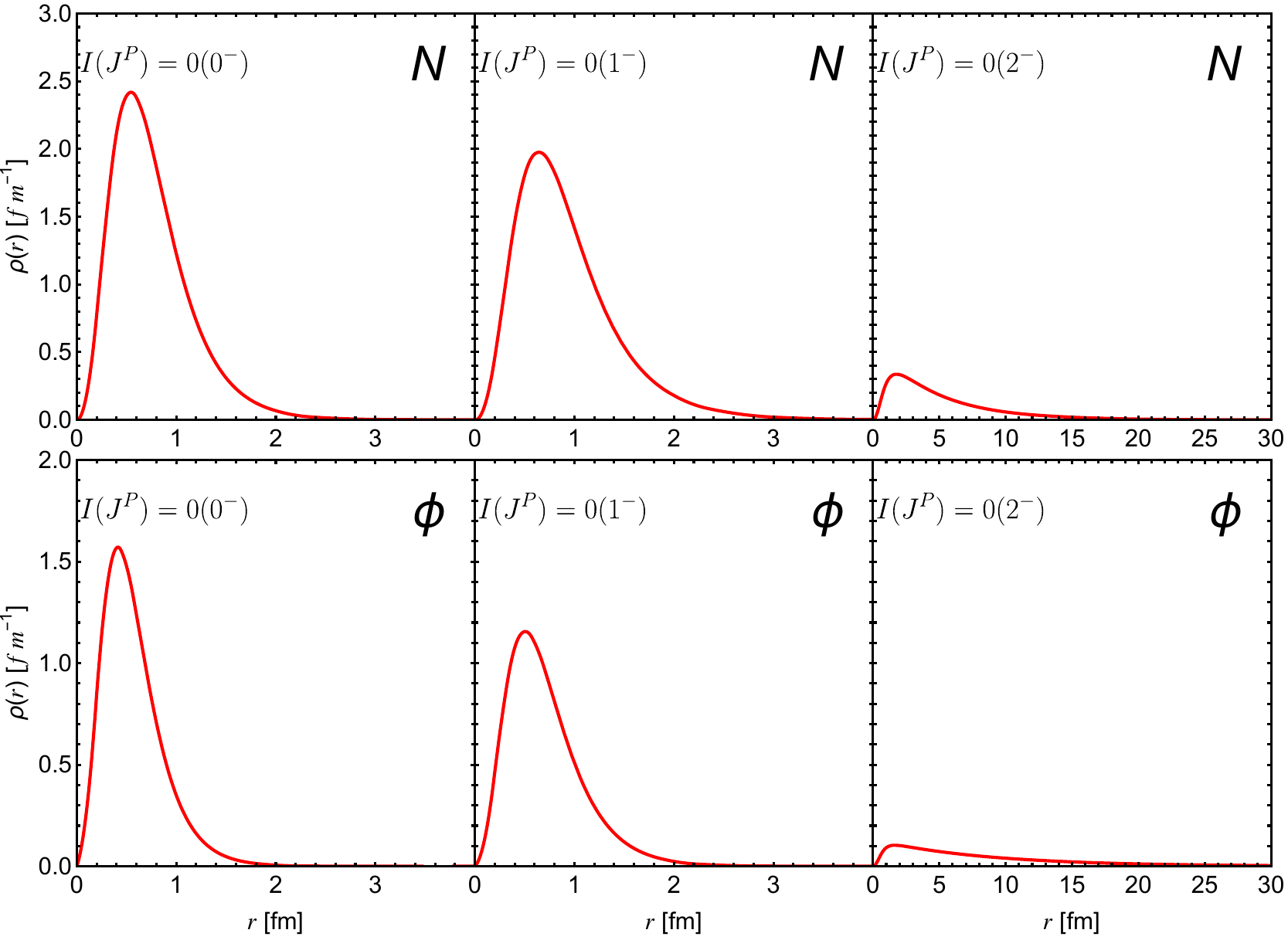} 
  \clabel[DtrA]{\caption{\label{fig:DtrA}The one-body density distributions $\rho(r)$ (in units of $fm^{-1}$) for nucleons and $\phi$ meson in the $\phi$NN states within Model A. The top subfigures show the one-body density distributions of nucleons, while the bottom subfigures show the one-body density distributions of the $\phi$ meson.}}
    \setlength{\belowdisplayskip}{1pt}
\end{figure*}

The $NN$(${}^3S_1$) channel features a well-known shallow bound state, the deuteron ($d$), with a binding energy of $2.41\ \mathrm{MeV}$. The relatively deeper binding energy observed in our calculations, compared to the experimental value, arises from our use of lattice hadron mass results to ensure consistency with the HAL QCD quarkonium-nucleon interactions. The ${}^1S_0$ channel exhibits an extremely large absolute scattering length, which is attributed to the presence of a nearby virtual state.

In the $\phi N$(${}^4S_{3/2}$) channel, no bound state is formed. However, the ${ }^2 S_{1 / 2}$ channel exhibits significantly different outcomes between the two models. Model A predicts a deeply bound state \change{$\tilde{N}$} with a binding energy of approximately 18 $\mathrm{MeV}$, whereas Model B shows no bound state but suggests strong attraction, as evidenced by a large absolute scattering length.

In the charmonium sector, no bound states are observed, and the absolute scattering lengths are relatively small. The negative scattering lengths indicate attractive interactions. Our calculated scattering parameters agree with those reported in Refs.~\cite{Lyu:2022imf,Lyu:2024ttm}. Notably, the spin-dependent interaction in charmonium-nucleon systems is weak, resulting in nearly identical scattering lengths and effective ranges for $J/\psi N$(${}^4S_{3/2}$) and $J/\psi N$(${}^2S_{1/2}$). This suggests that the spin interaction is significantly weaker than the central force in both charmonium-nucleon and $\phi N$ interactions. In Model B, the spin interaction is inversely proportional to mass, which aligns well with these findings.

\section{$\phi$ NN system}~\label{sec:results_phiNN}

\subsection{$\phi NN$ states in Model A}\label{subsec:A}

The eigenstates of the Model A complex-scaled Hamiltonian are shown in Fig.~\ref{fig:phiNNA}. We use blue, orange, and green markers to represent the results from three different complex scaling angles: $0^\circ ,9^\circ ,12^\circ$. We obtain discrete scattering states due to the finite basis employed in our calculations. The dashed lines represent the theoretical continuum scattering states for the respective angles.
 Each bound state remains at the same position regardless of the complex scaling angle. The binding energies and rms radii of bound states are provided in Table~\ref{tab:bound}.

 \begin{table*}[] 
    \renewcommand{\arraystretch}{1.4}
    \centering
    \caption{\label{tab:bound} The binding energies and rms radii of the ${\phi NN}$ bound states in two different models. The ${E_{B}}$ represent the binding energies relative to the corresponding lowest two-body thresholds, with the values in parentheses referring to the binding energies relative to the three-body threshold. In the column leading by ``2-body thresh.", the two-body thresholds is listed in ascending order. For comparison, the binding energies and the lowest two-body thresholds of $\phi NN$ states obtained from Ref.~\cite{Filikhin:2024avj} are presented.}
   \begin{tabular*}{\hsize}{@{}@{\extracolsep{\fill}}l|cccc|cccc|cc@{}}
\hline\hline
 \multirow{2}{*}{ ${I (J^{P})}$ } & \multicolumn{4}{c|}{Model A}  & \multicolumn{4}{c|}{Model B} & \multicolumn{2}{c}{Ref.~\cite{Filikhin:2024avj} } \\
       & 2-body thresh. & ${E_{B}}$ [MeV]  &$r^{{rms}}_{{N N}}$ [fm]& $r^{{rms}}_{{\phi N}}$ [fm]  & 2-body thresh. & ${E_{B}}$ [MeV]  &$r^{{rms}}_{N N}$  [fm]& $r^{{rms}}_{\phi N}$ [fm]  & 2-body thresh. & ${E_{B}}$ [MeV]  \\
     \hline 
       $0(0^-)$ & $\change{\tilde{N}}N,d\phi$ & $34.0(52.4)$ & $1.55$ & $1.26$ & $ d\phi$ & $0.95(3.35)$ & $3.10$ & $4.87$  & $ \tilde{N}N,d\phi$ & $36.5(64.1)$\\ 
       $0(1^-)$  & $ \change{\tilde{N}}N,d\phi$ & $15.2(33.6)$ & $1.81$ & $1.58$  & $ d\phi$ & $0.52(2.93)$ & $3.28$ & $6.21$  & $ d\phi$ & $12.7(14.9)$\\
       $1(1^-)$  & $ \change{\tilde{N}}N$ & ... & ... & ...  & ... & ... & ... & ... & ... & $...(5.47)$ \\
       $0(2^-)$  & $ d\phi$ & $0.04(2.45)$ & $3.68$ & $17.9$& $ d\phi$ & $0.04(2.45)$ & $3.68$ & $17.9$  & $ d\phi$ & $...(...)$\\

     \hline \hline
   \end{tabular*}
\end{table*}

In the $0(0^-)$ $\phi NN$ system, there are two two-body thresholds, $\tilde{N}N $ and $d\phi$, lie below the three-body threshold. A deep bound state is identified in the top-left panel of Fig.~\ref{fig:phiNNA}, with a binding energy of ${52.36 \ \mathrm{MeV}}$ relative to the three-body threshold and 34.0 MeV with respective to the $\change{\tilde{N}}N$ threshold. Among the three bound states listed in Table~\ref{tab:bound}, the $0(0^-)$ system is the most deeply bound because it involves only the strong $NN(^3S_1)$ and $\phi N(^2S_{1/2})$ interactions. Due to the tight binding, the $r^{{rms}}_{N N}=1.55\ \mathrm{fm}$ is significantly smaller than that of the deuteron, and the $r^{{rms}}_{\phi N}$ is the smallest among the three bound states in Table~\ref{tab:bound}. The $r^{{rms}}_{\phi N}$ increases monotonically with binding energy, making it useful for assessing the tightness of few-body bound states.  Notably, the $r^{{rms}}_{\phi N}$ is comparable to that of the two-body $\phi N$ bound state \change{$\tilde{N}$}. Since the $N N$ pairs are identical particles, the $r^{{rms}}_{{\phi N}}$ inherently includes contributions from both nucleons. \change{This can be addressed by considering the one-body density distributions of $\phi$. As shown in Fig.~\ref{fig:DtrA}, we find that the distribution of $\phi$ is peaked at approximately $0.5~\mathrm{fm}$, which is significantly smaller than the rms radius of the $\phi N({}^2S_{1/2})$ two-body bound states, $1.27~\mathrm{fm}$. This suggests a tighter configuration of the three-body bound states compared to the $\phi N$ two-body bound states.}

The $0(1^-)$ channel shares the same two-body thresholds, $\change{\tilde{N}}N $ and $d\phi$, with the $0(0^-)$ channel. The pair interaction between two nucleons involves only the $NN(^3S_1)$ channel, similar to the $0(0^-)$ $\phi NN$ system. The $0(1^-)$ channel differs by incorporating the mixing of $\phi N(^4S_{3/2})$ and $\phi N(^2S_{1/2})$ spin components. The overlap coefficients for these channels are provided in Eq.~\eqref{eq:bases_2}, where the strongly attractive component of $\phi N(^2S_{1/2})$ exhibits dominant overlap with $NN({}^3S_1)$. We find a bound state in the $0(1^-)$ channel with a binding energy of ${33.6 \ \mathrm{MeV}}$ relative to $\phi NN$ threshold and ${15.2 \ \mathrm{MeV}}$ relative to the lowest two-body threshold $\change{\tilde{N}} N $. The $\phi N(^4S_{3/2})$ component disfavors a bound state, resulting in a weaker interaction for the $S=1$, $I=0$ $\phi NN$ system compared to the $S=0$, $I=0$ case. The corresponding $r^{{rms}}$ are shown in Table~\ref{tab:bound}, where each $r^{{rms}}$ is slightly larger than its $0(0^-)$ partner. \change{Additionally, the distributions of $N$ and $\phi$ in the $0(1^-)$ channel are more dispersed compared to those in the $0(0^-)$ channel, as shown in Fig.~\ref{fig:DtrA}.}

For the $1(1^-)$ channel, we do not observe any bound states or resonances, as illustrated in Fig.~\ref{fig:phiNNA}. This absence is attributed to the smaller overlap matrix element between the strongly attractive component $\phi N(^2S_{1/2})$ and $NN({}^1S_0)$, compared to that in the $0(1^-)$ channel. From the perspective of discrete wave functions, the interaction between the $\phi N(^2S_{1/2})$ pair and the spectator $N$ in the $1(1^-)$ channel is weaker by a factor of $1/\sqrt{2}$ compared to that in the $0(1^-)$ sector (see Eqs.~\eqref{eq:bases_2} and \eqref{eq:bases_4}), which disfavors the formation of three-body bound states.
 Additionally, the $NN$ interaction in the $1(1^-)$ $\phi NN$ system corresponds to the $NN({}^1S_0)$ virtual state, which is weaker than the $NN({}^3S_1)$ interaction associated with the deuteron. By adjusting the interaction parameters, we find that the $1(1^-)$ state behaves as a virtual state relative to the $\change{\tilde{N}}N$ threshold, while being a bound state relative to the three-body threshold. Thus, the existence of few-body bound states requires not only a strong interaction but also a competition with sub-channels.
\clabel[reply4]{It is worth noting that the quantum numbers of $\tilde{N}$ align with those of nucleon excited states $N^*$, such as $N^*(1535)$. Therefore, when the conversion potential is taken into account, the three-body bound states of $\phi NN$ become quasi-bound states that can decay into the $N^* N$ channel. However, given that the dominant Fock state of $\tilde{N}$ contains an $s\bar{s}$ component, it is expected that the decay width would be relatively small compared to the rearrangement width, if the latter exists.} 
 \subsection{$\phi NN$ states in Model B}
\begin{figure*}[]
\centering
  \includegraphics[width=0.83\textwidth]{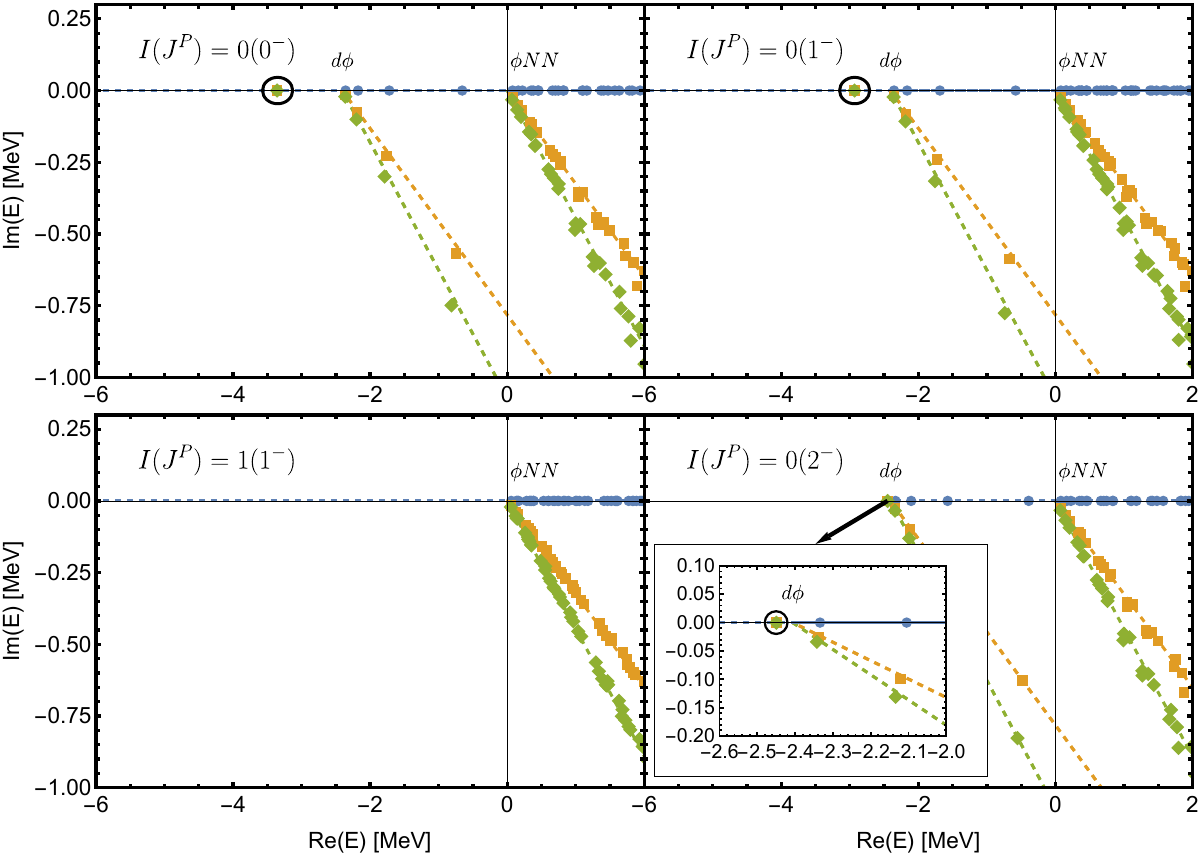} 
  \caption{\label{fig:phiNNB}The complex energy eigenvalues of the $\phi NN$ states in Model B with varying $\theta$ in the CSM. The dashed lines represent the continuum lines rotating along ${Arg}(E)=-2\theta$. The bound states do not shift as $\theta$ changes and are marked out by the black circles.}
    \setlength{\belowdisplayskip}{1pt}
\end{figure*}

\begin{figure*}[]
\centering
  \includegraphics[width=0.83\textwidth]{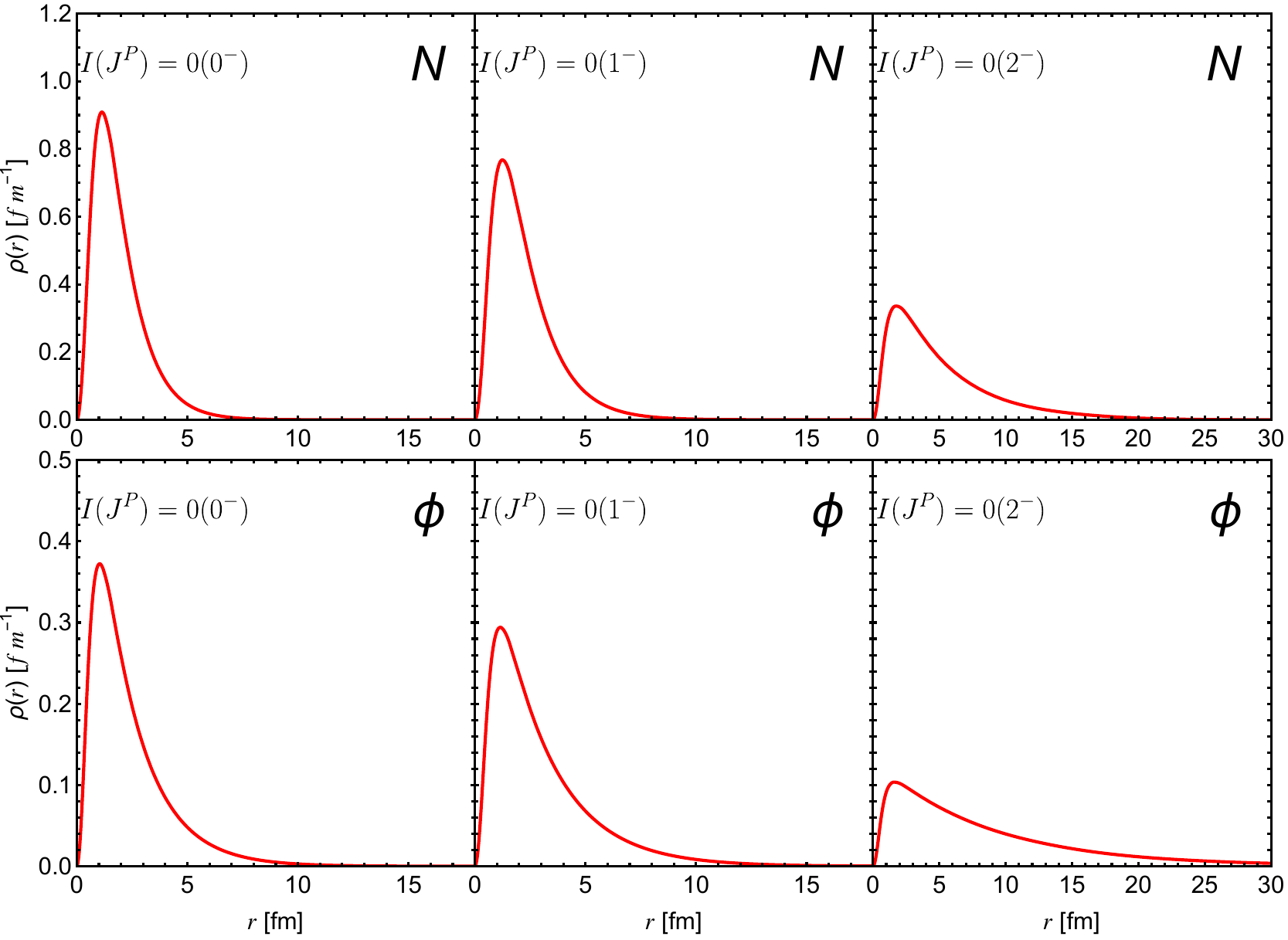} 
  \clabel[DtrB]{\caption{\label{fig:DtrB}The one-body density distributions $\rho(r)$ (in units of $fm^{-1}$) for nucleons and $\phi$ meson in the $\phi$NN states within Model B. The top subfigures show the one-body density distributions of nucleons, while the bottom subfigures show the one-body density distributions of the $\phi$ meson.}}
    \setlength{\belowdisplayskip}{1pt}
\end{figure*}

We note that Ref.~\cite{Filikhin:2024avj} employs an average potential to treat the $\phi N(^2S_{1/2})$ and $\phi N(^4S_{3/2})$ channels, rather than performing a complete coupled-channel calculation. The average $\phi N$ potentials for $0(1^-)$ and ${1(1^-)}$ three-body systems are given by:
\begin{equation}
\begin{aligned}
&\bar{V}_{\phi N}^{0(1^-)}=\frac{2}{3}V_{\phi N(^2S_{1/2})}+\frac{1}{3}V_{\phi N(^4S_{3/2})},\\
&\bar{V}_{\phi N}^{1(1^-)}=\frac{1}{3}V_{\phi N(^2S_{1/2})}+\frac{2}{3}V_{\phi N(^4S_{3/2})}.
\end{aligned},
\end{equation}
Apparently, this average potential approach is merely an assumption and lacks dynamical detail. What is more important, the averaged $\phi N$ interaction would prevent the existence of the $\change{\tilde{N}}$ two-body $\phi N(^2S_{1/2})$ bound states and consequently make the two-body scattering states $\change{\tilde{N}}N$  absent in the three body system, which would essentially affects the analyticity of the S-matrix. They obtained bound states in both the $0(1^-)$ and $1(1^-)$ channels, as shown in Table~\ref{tab:bound}. The binding energy of the $0(1^-)$ state is 12.7 $\mathrm{MeV}$ relative to the $d\phi$ two-body threshold and 14.9 $\mathrm{MeV}$ relative to the three-body threshold, whereas the binding energy of the $1(1^-)$ state is 5.47 $\mathrm{MeV}$ relative to the three-body threshold. Notably, the binding energy of the $0(1^-)$ state is significantly underestimated when compared to complete dynamical calculations. In the case of the $1(1^-)$ channel, the average potential approach results in a spurious bound state.

The spin-2 channel features a very shallow bound state with a binding energy of only ${0.04 \ \mathrm{MeV}}$ relative to the $d\phi$ threshold, as shown in the last column of Table~\ref{tab:bound}. This channel comprises only the $\phi N({{}^4S_{3/2}})$ component, depending exclusively on HAL QCD $\phi N({}^4S_{3/2})$ potential. The $r^{{rms}}_{N N}$ of this state is similar to that of the deuteron, while the $r^{{rms}}_{\phi N}$ is approximately ${17.9 \  \mathrm{fm}}$.
The structure of this state resembles a $\phi-d$ atom, with the $\phi$ meson surrounding the deuteron. It is important to note that our results are based on the original parameters at \( m_\pi = 146.4\ \mathrm{MeV} \) from HAL QCD~\cite{Lyu:2022imf}. In contrast, Ref.~\cite{Filikhin:2024avj} reports an inability to find a bound state in this channel. This discrepancy likely arises from their substitution of the pion mass in \eqref{eq:phiN3} with the physical pion mass. The two-pion exchange term includes an \( m_\pi^4 \) factor, which amplifies the effects of small changes in the pion mass.  The present parametrized two-pion exchange term, $e^{-2 m_\pi r}/r^2$, contributes significantly to the short-range interaction, particularly in the range \( 0.5\ \text{fm} < r < 1\ \text{fm} \). However, one could construct different mechanisms (with different pion mass dependence) to model the short-range interaction.  
%In principle,  the details of the short-range interaction should not be %exclusively determined by the low-energy observables.  
Consequently, it is difficult to believe that a straightforward substitution of $m_\pi$ exactly captures the true pion mass dependence in the short-range regime. 

In Model A, relatively deep bound states exist for the $0(1^-)$ and $1(1^-)$ systems. Even when considering the uncertainties in the parameters of the $\phi N({{}^4S_{3/2}})$ potential listed in Table~\ref{tab:paraphiN}, their existence is unlikely to be affected, only with the potential shifting of their binding energies. Since this work focuses primarily on qualitative results, we will not delve further into the uncertainties of their binding energies, as their existence is established.  However, for the $0(2^-)$ $\phi NN$ system, its existence could be affected by uncertainties in the potential parameters. As the interactions for the $0(2^-)$ system are identical in both Model A and Model B, we will investigate its uncertainties in the following section.

The eigenstates of the complex-scaled Hamiltonian for Model B are illustrated in Fig.~\ref{fig:phiNNB}. The three-body bound states in the  nominal values of the parameters in Tables~\ref{tab:paraphiN} and \ref{tab:parajpsiN} are summarized in Table~\ref{tab:mass}. In Model B, there is no $\phi N$ two-body bound state. The difference between the $\phi N({}^4S_{3/2})$ and $\phi N({}^2S_{1/2})$ interactions is relatively minor. The $\phi N({}^2S_{1/2})$ interaction exhibits slightly stronger attraction than the $\phi N({}^4S_{3/2})$. The binding energy order of the three isoscalar $\phi N N$ depends on the components of the $(\phi N)_{{1\over2},{1\over2}}N$, as detailed in Appendix~\ref{AppendixB}, which represents the most attractive spin channel. Specifically, the greater the contribution of $(\phi N)_{{1\over2},{1\over2}}N$ there is, the deeper the bound states become. Thus, we have the following order 
\begin{equation}
    E_b^{0(2^-)}<E_b^{0(1^-)}<E_b^{0(0^-)}.
\end{equation}
The presence of a shallow $0(2^-)$ $\phi NN$ bound state suggests that similar shallow bound states should also exist for the $0(0^-)$ and $0(1^-)$ channels. Specifically, the $0(0^-)$ and $0(1^-)$ bound states are situated near the $d\phi$ threshold, with binding energies of ${0.95 \ \mathrm{MeV}}$ and ${0.52\ \mathrm{MeV}}$, respectively. \change{The radii of $N$ and $\phi$ in these bound states are significantly more extended compared to the range of the two-body force within 1 fm, consistent with the shallow binding energy in each channel, as shown in Fig.~\ref{fig:DtrB}.} Consequently, the bound states in each channel can be characterized as $\phi$-d two-body molecular state. This characterization is further supported by the observation that $r^{{rms}}_{NN}\approx r^{{rms}}_{{d}}$. 

\begin{figure}[htbp]
\centering
  \includegraphics[width=0.48\textwidth]{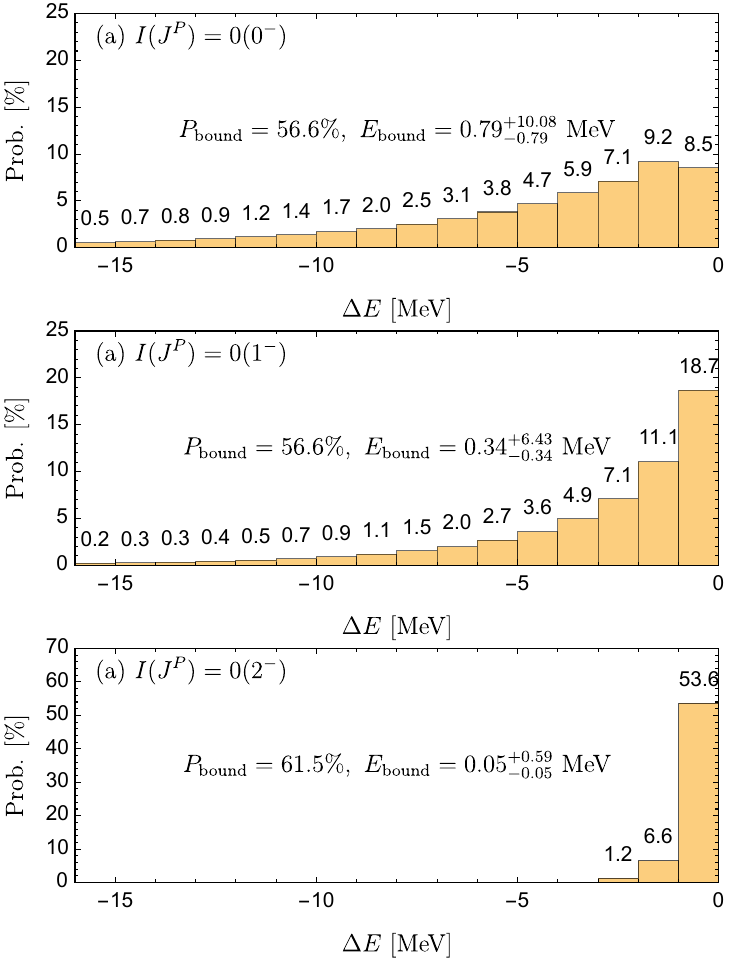} 
  \caption{\label{fig:Prob}The distribution of the $\phi NN$ binding energies in the $0(0^-)$, $0(1^-)$, and $0(2^-)$ channels accounting for parameter uncertainties in Model B. The numbers on each bar represent the percentage of bound states within 1 $\mathrm{MeV}$ energy intervals.}
    \setlength{\belowdisplayskip}{1pt}
\end{figure}

Due to the relatively weak binding energies in $\phi NN$ systems, uncertainties cannot be ignored in the analysis. Therefore, we extracted the uncertainties from the HAL QCD potential function for both the $\phi N$ and $J/\psi N$ interactions. For each parameter $\alpha_1, \alpha_2, \alpha_3, \beta_1, \beta_2, \beta_3$ in $\phi N({}^4S_{3/2})$ interactions, we selected three points: the central value, the central value plus one standard deviation ($\sigma$), and the central value minus one standard deviation ($\sigma$), resulting in a total of $3^6$ parameter sets. 
For the spin-dependent interaction of \( \phi N \), determined from \( J/\psi N \) interactions, we only account for the uncertainties in the linear coefficients and use perturbation theory to propagate these uncertainties. This approach is justified as the spin-dependent interaction is relatively weak compared to the central interaction. Specifically, for the $J/\psi N({}^4S_{3/2})$ and $J/\psi N({}^2S_{1/2})$ interactions, we varied each parameter $\alpha_1, \alpha_2, \alpha_3$ individually within the $\phi N$ interaction parameter space, generating 6 parameters across two channels. By comparing the changes in binding energy of bound states with and without altering the $J/\psi N$ interaction parameters, we assessed the impact of $J/\psi N$ interactions on the binding energies and existence of bound states in the $\phi N$ parameter space. This process ultimately resulted in a dataset of binding energies comprising $3^{12}$ parameter sets. We introduced a binding probability of three-body bound state $P_{\rm bound}$ across the parameter space. Its distribution is shown in Fig.~\ref{fig:Prob}. 
We find that, due to the uncertainties in the HAL QCD extraction of the $\phi N$ and $J/\psi N$ interactions, the existence of a three-body bound state is not certain. Only approximately $60\%$ of the parameter sets support the formation of bound states in $\phi NN$ systems. We also use the 68\% probability (1 $\sigma$ significance) to estimate the uncertainty of the binding energies, which is summarized in Fig.~\ref{fig:Prob}. Under the 1 $\sigma$ significance, the $0(0^-)$ $\phi NN$ state varies from an unbound state to the bound state with a binding energy 10.87 $\mathrm{MeV}$. For the $0(1^-)$ $\phi NN$ state, it varies from an unbound state to the bound state with a binding energy 6.77 $\mathrm{MeV}$. The $0(2^-)$ $\phi NN$ state has the smallest binding energy range, varying from an unbound state to the bound state with a binding energy 0.64 $\mathrm{MeV}$.

%  \begin{table*}[htbp] 
%     \renewcommand{\arraystretch}{1.4}
%     \centering
%     \caption{\label{tab:Prob} The probability of forming bound states ($P_{bound}$) and the range of binding energies ($E_{Theo.}$) for the $\phi NN$ bound states in Model B are presented. The $Prob$ is determined based on the 1-sigma lower limit.}
% \begin{tabular*}{\hsize}{@{}@{\extracolsep{\fill}}lccc@{}}

% \hline\hline
%   $I(J^P)$ &$P_{bound}$ & $E_{{Theo.}}[\mathrm{MeV}]$    \\
%  \hline 
%  $0(0^-)$ & $56.6\%$  &  $0\sim 10.87$  \\
%  $0(1^-)$ & $56.6\%$  &  $0\sim 6.77$  \\
%  $0(2^-)$ & $61.5\%$  &  $0\sim 0.05$  \\
% \hline \hline
% \end{tabular*}
% \end{table*}

\section{$J/\psi NN$ and $\eta_c NN$ states}\label{sec:charmonium}
\begin{figure*}[htbp]
\centering
  \includegraphics[width=0.83\textwidth]{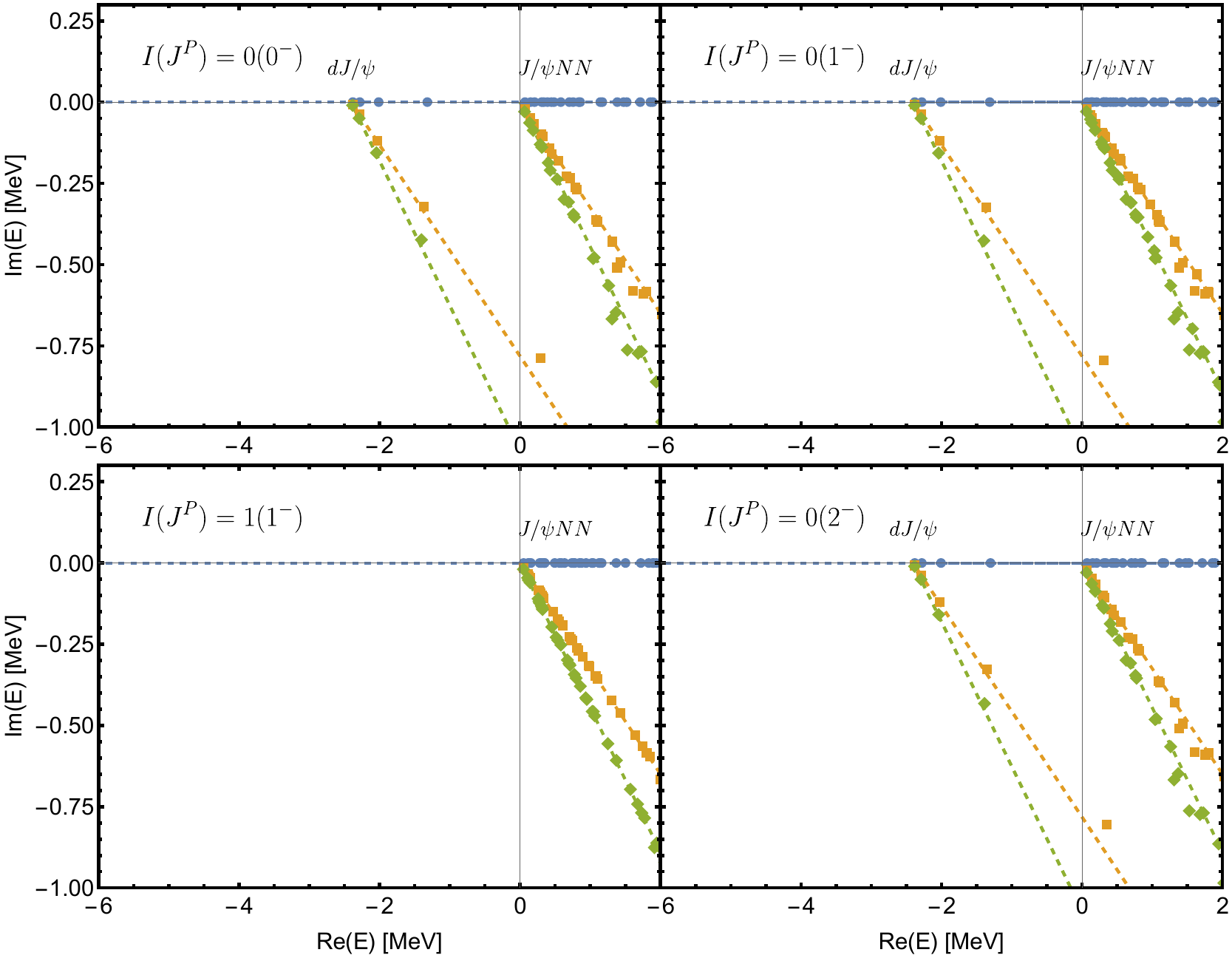} 
  \caption{\label{fig:S1}The complex energy eigenvalues of the ${J/\psi NN}$ states with varying $\theta$ in the CSM. The dashed lines represent the continuum lines rotating along ${Arg}(E)=-2\theta$. No bound state or resonance exists.}
    \setlength{\belowdisplayskip}{1pt}
\end{figure*}

\begin{figure*}[htbp]
\centering
  \includegraphics[width=0.83\textwidth]{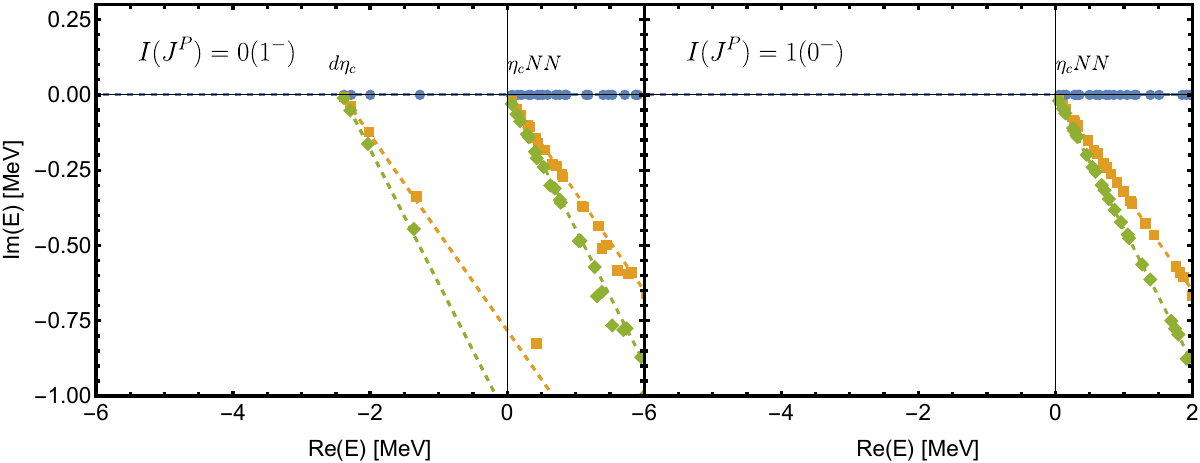} 
  \caption{\label{fig:S2}The complex energy eigenvalues of the ${\eta_c NN}$ states with varying $\theta$ in the CSM. The dashed lines represent the continuum lines rotating along ${Arg}(E)=-2\theta$. No bound state or resonance exists.}
    \setlength{\belowdisplayskip}{1pt}
\end{figure*}

Our results for $J/\psi N N$ and ${\eta_c N N}$ based on the GEM are shown in Figs.~\ref{fig:S1} and~\ref{fig:S2}, respectively. Although both $J/\psi N$ and $\eta_cN$ interactions are attractive, we find that neither bound states nor resonances exist in  $J/\psi N N$ and ${\eta_c N N}$ systems. This conclusion is consistent with previous GEM calculations~\cite{Yokota:2013sfa}, which were based on earlier quenched lattice QCD (LQCD) results~\cite{Kawanai:2010ev}.


We also estimate the requirement of the two-body scattering lengths of $J/\psi N$ or $\eta_c  N$ in order to form the three-body bound states. Using HAL QCD Gaussian-type potentials in Eq.~\ref{eq:jpsietacN} with an extra adjustable factor, we determine that for a three-body bound state to exist, the two-body scattering length must be less than approximately ${-1\ \mathrm{fm}}$. This finding is consistent with that using the Gaussian type and Yukawa type potential in Ref.~\cite{Yokota:2013sfa}. Therefore, the two-body scattering length serves as a crucial criterion for identifying three-body bound states.

\begin{figure}[htbp]
\centering
  \includegraphics[width=0.45\textwidth]{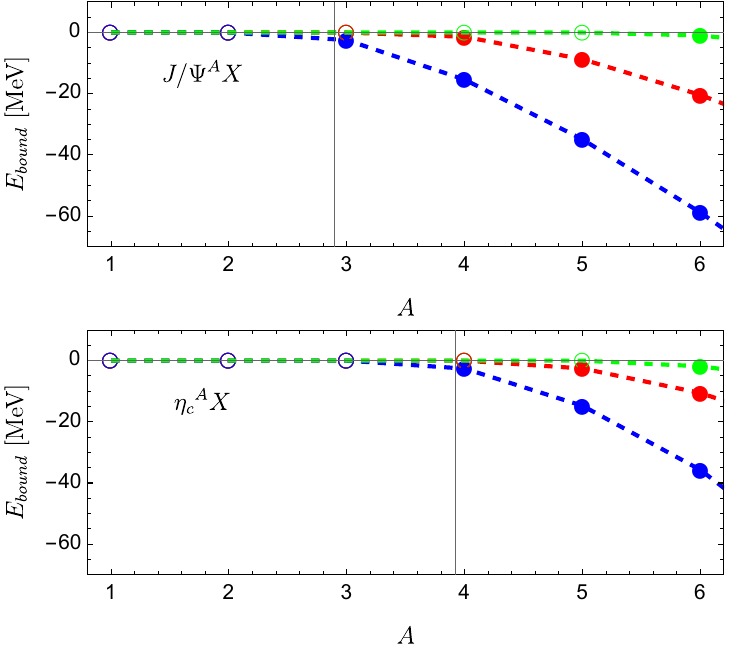} 
  \caption{\label{fig:EA}The binding energies (in $\mathrm{MeV}$) of the $(c\bar{c})$-nuclei systems vary with nuclear mass number \( A \). Solid circles represent existing bound states, while hollow circles denote the absence of bound states. The red, green, and blue lines (or markers) correspond to results obtained using the central values, upper limits, and lower limits of the parameters, respectively, to account for uncertainties.}
    \setlength{\belowdisplayskip}{1pt}
\end{figure}

With the suppression of the spin-dependent $J/\psi N$ interaction and the spin-independence of the $\eta_c N$ interaction, we can roughly estimate the interaction between charmonium and a nucleus with more than two nucleons. Since both the $J/\psi N$ and $\eta_c N$ interactions are expected to be isospin-independent and largely insensitive to spin, we can use the potential \( V_{(c\bar{c})N} \) (taking the spin-averaged value for the $J/\psi N$ system) to approximate the interaction \( V_{(c\bar{c}){}^A X} \):  
\[
V_{(c\bar{c}){}^A X} = A \times V_{(c\bar{c})N},
\]
where \( A \) is the mass number of the nucleus. Using this approach, we can estimate the minimum \( A \) required for the $(c\bar{c})$ to bind to a nucleus ${}^A X$. The binding energy is shown in Fig.~\ref{fig:EA}. It is evident that, with nominal parameters, the lightest nucleus capable of binding $(c\bar{c})$ has a mass number \( A = 4 \), such as ${}^4 \mathrm{He}$. Considering uncertainties, there is still a significant probability that nuclei with \( A = 3 \), such as ${}^3 \mathrm{H}$ or ${}^3 \mathrm{He}$, could also bind $J/\psi$. For $\eta_c$, a nucleus with mass number \( A = 5 \) can bind $\eta_c$ with nominal parameters, and there are substantial possibilities to form bound states with \( A = 4 \).

\section{Summary and Discussion}~\label{sec:sum}
We investigate the ground states of the ${\phi NN}$ and ${J/\psi NN}$ systems with isospin-spin-parity combinations $I(J^{P})=0(0^-), 0(1^-), 1(1^-), 0(2^-)$, as well as the ${\eta_c NN}$ system with $I(J^{P})=1(0^-), 0(1^-)$, based on HAL QCD interactions. The Gaussian expansion method is employed to solve the complex-scaled Sch\"ordinger equation. 

For the $\phi NN$ system, the lattice QCD $\phi N\left({ }^2 S_{1 / 2}\right)$ interaction is not available. In this work, we explore the interaction using two models: Model A, which combines the  $\phi p$ correlation function analysis and HAL QCD results, and  Model B, which assumes that the spin-spin interactions for  $J/\psi N$ and $\phi N$ systems are inversely proportional to their masses. 
Model A predicts a stronger $\phi N({}^2 S_{1/2})$ interaction, permitting a two-body bound state, while Model B indicates that the interaction is attractive but not strong enough to support a bound state. Both models, however, predict bound states for the $I(J^P) = 0(0^-)$ and $0(1^-)$ $\phi NN$ systems. In Model A, these states are deeply bound with binding energies exceeding 15 MeV and remain existent when considering parameter uncertainties. In contrast, these states are very loosely bound in Model B, with binding energies below 1 MeV and existent probability of about 60\% when parameter uncertainties are considered. In both models, there exist very loosely bound $I(J^P) = 0(2^-)$ three-body states. However its existence is sensitive to parameter uncertainties. Additionally, we find no bound states or resonances in the isovector $I(J^P) = 1(1^-)$ $\phi NN$ system in either model.

We also calculate the root-mean-square (rms) radii, $r^{rms}$, \change{as well as the one-body density distributions, $\rho(r)$}, of these states to further elucidate their spatial configurations. In Model A, the bound states with $I(J^{P})=0(0^-)$ and $0(1^-)$ are tightly bound, leading to significantly smaller $r^{rms}_{NN}$ compared to $r^{rms}_{d}$. In contrast, in Model B, the bound states with $I(J^{P})=0(0^-)$ and $0(1^-)$ are shallow, resembling a $\phi$-d atom, similar to the $0(2^-)$ bound state. However, due to the indistinguishability of nucleons, the spatial structure of the tightly bound states cannot be fully determined using $r^{rms}_{\phi N}$ alone. \change{We utilize the one-body density distributions to further determine the tightness of the three-body bound states.}

We also investigate the ground state in $J/\psi NN$ and $\eta_c NN$ three-body systems using the newly proposed HAL QCD potential. For $J/\psi NN$, we consider $I(J^{P})=0(0^-),0(1^-),1(1^-),0(2^-)$, and for $\eta_c NN$, we consider  $0(1^-),1(0^-)$. No bound states or resonances are found in these systems, which can be attributed to the small absolute value of the two-body scattering length in the $(c\bar{c}) N$ interactions. 

The criterion for the existence of a three-body bound state is that the two-body $(c\bar{c}) N$ scattering length is approximately less than ${-1\ \mathrm{fm}}$, which lies outside the range of most previous experimental and theoretical approaches.
%Based on previous experimental and theoretical studies, satisfying this %criterion is challenged. 
Furthermore, We also estimate the interaction between charmonium and nuclei, concluding that the $J/\psi$ or $\eta_c$ is likely to form bound states with ${}^3\mathrm{H}$, ${}^3\mathrm{He}$, ${}^4\mathrm{He}$, and heavier nuclei. Therefore, we suggest exploring the possibility of these bound states in the future.

\begin{appendix}

\section{Overlap of spin-flavor wave functions}~\label{AppendixB}
The overlap matrix of the spin-flavor wave functions is presented below with the bases $[(NN)_{1,0}\phi]$ for the $I=0$ system ($[(NN)_{0,1}\phi]$ for the $I=1$ system), $[(\phi N)_{\frac{1}{2},\frac{1}{2}}N]$, $[(\phi N)_{\frac{3}{2},\frac{1}{2}}N]$, $[N(\phi N)_{\frac{1}{2},\frac{1}{2}}]$, and $[N(\phi N)_{\frac{3}{2},\frac{1}{2}}]$:
% 第一个方程：S = 0, I = 0
\begin{equation}\label{eq:bases_1}
\begin{pmatrix}
        1 & \frac{1}{\sqrt{3}} & -\sqrt{\frac{2}{3}} & -\frac{1}{\sqrt{3}} & \sqrt{\frac{2}{3}} \\
        \frac{1}{\sqrt{3}} & 1 & 0 & \frac{1}{3} & \frac{2 \sqrt{2}}{3} \\
        -\sqrt{\frac{2}{3}} & 0 & 1 & \frac{2 \sqrt{2}}{3} & -\frac{1}{3} \\
        -\frac{1}{\sqrt{3}} & \frac{1}{3} & \frac{2 \sqrt{2}}{3} & 1 & 0 \\
        \sqrt{\frac{2}{3}} & \frac{2 \sqrt{2}}{3} & -\frac{1}{3} & 0 & 1
    \end{pmatrix}
\end{equation}
for the $S=0,I=0$ system;
% 第二个方程：S = 1, I = 0
\begin{equation}\label{eq:bases_2}
    \begin{pmatrix}
        1 & -\sqrt{\frac{2}{3}} & -\frac{1}{\sqrt{3}} & \sqrt{\frac{2}{3}} & \frac{1}{\sqrt{3}} \\
        -\sqrt{\frac{2}{3}} & 1 & 0 & -\frac{1}{3} & -\frac{2 \sqrt{2}}{3} \\
        -\frac{1}{\sqrt{3}} & 0 & 1 & -\frac{2 \sqrt{2}}{3} & \frac{1}{3} \\
        \sqrt{\frac{2}{3}} & -\frac{1}{3} & -\frac{2 \sqrt{2}}{3} & 1 & 0 \\
        \frac{1}{\sqrt{3}} & -\frac{2 \sqrt{2}}{3} & \frac{1}{3} & 0 & 1
    \end{pmatrix}
\end{equation}
for the $S=1, I=0$ system; 

% 第三个方程：S = 2, I = 0
\begin{equation}\label{eq:bases_3}
        % S = 2, I = 0: & \\
    \begin{pmatrix}
            1 & 0 & -1 & 0 & 1 \\
            0 & 0 & 0 & 0 & 0 \\
            -1 & 0 & 1 & 0 & -1 \\
            0 & 0 & 0 & 0 & 0 \\
            1 & 0 & -1 & 0 & 1
    \end{pmatrix}
\end{equation}
for the $S=2,I=0$ system;

% 第四个方程：S = 1, I = 1
\begin{equation}\label{eq:bases_4}
    % S = 1, I = 1:& \\
    \begin{pmatrix}
        1 & \frac{1}{\sqrt{3}} & -\sqrt{\frac{2}{3}} & -\frac{1}{\sqrt{3}} & \sqrt{\frac{2}{3}} \\
        \frac{1}{\sqrt{3}} & 1 & 0 & \frac{1}{3} & \frac{2 \sqrt{2}}{3} \\
        -\sqrt{\frac{2}{3}} & 0 & 1 & \frac{2 \sqrt{2}}{3} & -\frac{1}{3} \\
        -\frac{1}{\sqrt{3}} & \frac{1}{3} & \frac{2 \sqrt{2}}{3} & 1 & 0 \\
        \sqrt{\frac{2}{3}} & \frac{2 \sqrt{2}}{3} & -\frac{1}{3} & 0 & 1
    \end{pmatrix}
\end{equation}
for the $S=1,I=1$ system.

\end{appendix}

\begin{acknowledgements}

We are grateful to Yan Lyu, Yan-Ke Chen, Jun-Zhang Wang and Zi-Yang Lin for the helpful discussions. 
This project was supported by the National Natural Science Foundation of China (Grant No. 12475137), and ERC NuclearTheory (Grant No. 885150). The computational resources were supported by High-performance Computing Platform of Peking University.

\end{acknowledgements}

\bibliography{Ref}

\end{document}